\documentclass[12pt]{iopart}

\usepackage{graphicx}  

\begin{document}

\newcommand{\ket}[1]{\ensuremath{\left|{#1}\right\rangle}}
\newcommand{\bra}[1]{\ensuremath{\left\langle{#1}\right|}}
\newcommand{\braket}[1]{\ensuremath{\left\langle{#1}\right\rangle}}

\title[Photodetecting propagating microwaves in circuit QED]{\center Photodetection of propagating quantum microwaves \\ in circuit QED}

\author{Guillermo Romero$^1$, Juan Jos\'e Garc\'{\i}a-Ripoll$^2$, \\ and Enrique Solano$^{3,4}$}

\address{$^1$Departamento de F\'{\i}sica, Universidad de Santiago de Chile, USACH, Casilla 307, Santiago 2, Chile}
  
\address{$^2$Instituto de F\'{\i}sica Fundamental, CSIC, Serrano 113-bis, 28006 Madrid, Spain}
  
\address{$^3$Departamento de Qu\'{\i}mica F\'{\i}sica, Universidad del Pa\'{\i}s Vasco - Euskal Herriko Unibertsitatea, Apdo. 644, 48080 Bilbao, Spain}
  
\address{$^4$IKERBASQUE, Basque Foundation for Science, Alameda Urquijo 36, 48011 Bilbao, Spain}
  
\ead{enrique$\!_{\huge\_}$solano@ehu.es}

\begin{abstract}
We develop the theory of a metamaterial composed of an array of discrete quantum absorbers inside a one-dimensional waveguide that implements a high-efficiency microwave photon detector. A basic design consists of a few metastable superconducting nanocircuits spread inside and coupled to a one-dimensional waveguide in a circuit QED setup. The arrival of a {\it propagating} quantum microwave field induces an irreversible change in the population of the internal levels of the absorbers, due to a selective absorption of photon excitations. This design is studied using a formal but simple quantum field theory, which allows us to evaluate the single-photon absorption efficiency for one and many absorber setups. As an example, we consider a particular design that combines a coplanar coaxial waveguide with superconducting phase qubits, a natural but not exclusive playground for experimental implementations. This work and a possible experimental realization may stimulate the possible arrival of "all-optical" quantum information processing with propagating quantum microwaves, where a microwave photodetector could play a key role.
\end{abstract}


\vspace{2pc}

\submitto{\PS}

\maketitle

\section{Introduction}
\label{sec:intro}

In a recent work~\cite{romero09} we suggested a possible implementation
of a photon detector that may also work as a photon counter in the
microwave regime. Our proposal builds on previous advances in the
field of quantum circuits in two
fronts. One is the development of artificial atoms and qubits~\cite{makhlin01,you05,schoelkopf08,clarke08}
for quantum computation and quantum information processing, using
quantized charge~\cite{bouchiat98,nakamura99, vion02,yamamoto03},
flux~\cite{mooij99,vanderwal00,chiorescu03,deppe08}, or
phase~\cite{martinis85,berkley03b,simmonds04,steffen06} degrees of
freedom. The other front is the efficient coupling of these elements
to microwave guides and cavities conforming the emergent field of circuit quantum electrodynamics (QED)~\cite{blais04,wallraff04,chiorescu04}. Without neglecting important advances in intracavity field physics in circuit QED, as we will continue to illustrate below, here we are interested in the physics of {\it propagating} quantum microwaves.

As we argued before~\cite{romero09}, in order to fully unleash the
power of quantum correlations in propagating microwave photonic fields, as may be generated by circuit QED setups, the implementation of efficient photon detectors and counters would be mostly welcomed. The existence of these detectors is implied by almost any sophisticated quantum protocol involving optical photons, be in their coherent interaction with matter~\cite{bouwmeester08} or purely all-optical devices~\cite{kok07}. It ranges from the characterization and reconstruction of nonclassical states of propagating light by quantum homodyne tomography~\cite{leonhardt97} to high-fidelity electron-shelving atomic qubit readout~\cite{leibfried03}. Both examples coming from quantum optics have shown to be influential in the novel field of circuit QED, with the first theoretical~\cite{mariantoni05} and experimental efforts~\cite{privatecommWMI} to measure relevant observables of propagating microwaves, and a recent proposal of mesoscopic shelving qubit readout~\cite{englert09}. In spite of these efforts, it will be very hard to overcome the necessity of photon detectors and counters when the emerging field of quantum microwaves will want to deal with local and remote interqubit/intercavity quantum communication, implementations of quantum cryptography, and other key advanced quantum information protocols~\cite{bouwmeester08,kok07}.

It should be thus no wonder that photon detection and counting become soon a central topic in the field of quantum circuits, where superconducting circuits interact with intracavity and propagating quantum microwaves. So far we have seen the
exchange of individual photons between superconducting qubits and
quantum resonators~\cite{majer07,hofheinz08,hofheinz09}, the resolution of photon number states in a superconducting circuit~\cite{schuster07}, the generation of propagating single photons~\cite{houck07}, the first theoretical efforts for detecting travelling photons~\cite{helmer09a,kumar09}, and the nonlinear effects that arise from the presence of a qubit in a resonator~\cite{fink08,bishop09}. We envision a rich dialogue between intracavity and intercavity physics in the microwave domain, see for example~\cite{mariantoni08,helmer09b,zhou08}, where matter and photonic qubits exchange quantum information in properly activated quantum networks for the sake of quantum information processing.

All efforts towards the implementation of a photodetector for propagating microwaves in circuit QED face a number of challenges, many of which are related to the specific nature of quantum circuits~\cite{romero09}. These are: i) Available cryogenic linear amplifiers are unable to resolve the few photon regime. ii) Free-space
cross-section between microwave fields and matter qubits are known to
be small. iii) The use of cavities to enhance the coupling introduces
additional problems, such as the frequency mode matching and the
compromise between high-Q and high reflectivity. iv) The impossibility
of performing continuous measurement without
backaction~\cite{helmer09a}, which leads to the problem of
synchronizing the detection process with the arrival of the measured
field. In a wide sense, the photodetection device has to be passive, being activated irreversibly by the arriving microwave signal. Otherwise, the advanced information that a photon is approaching turns itself into a photodetection device.

\begin{figure}[t]
\centering
\includegraphics[width=0.37\linewidth]{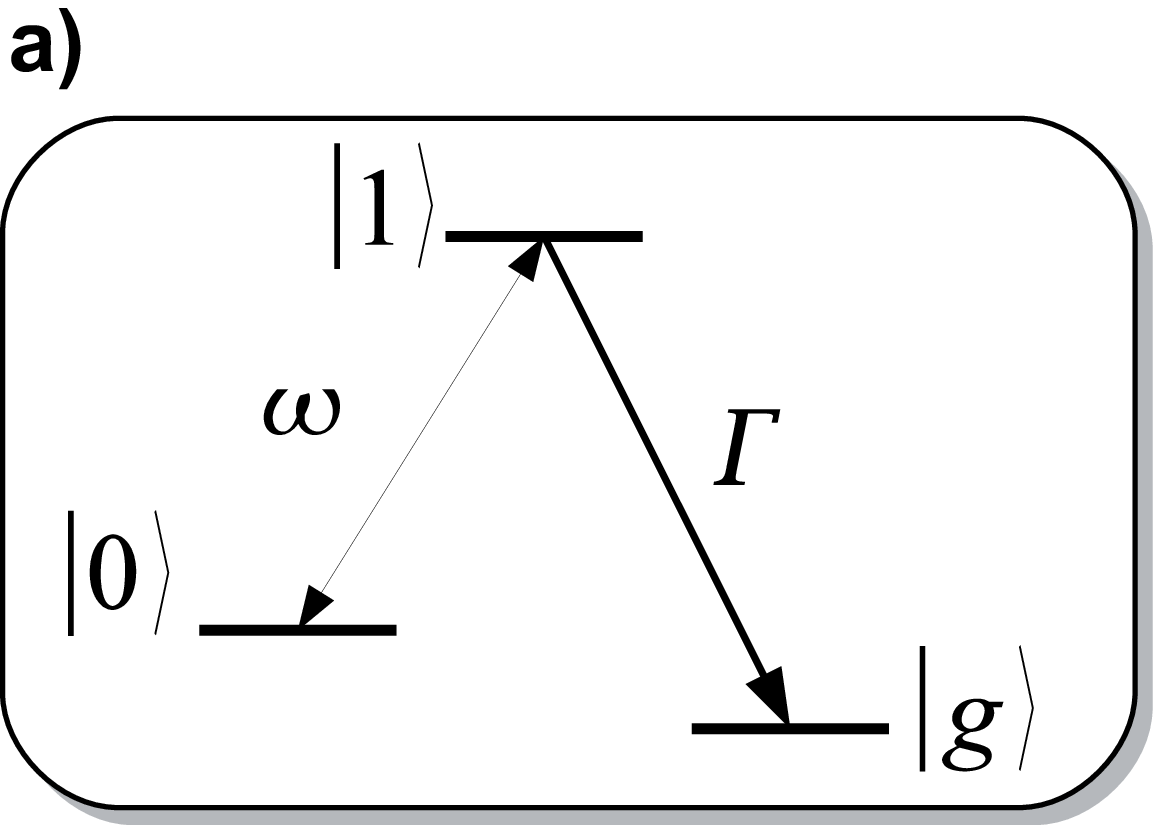}\,\,\,\,\includegraphics[width=0.5\linewidth]{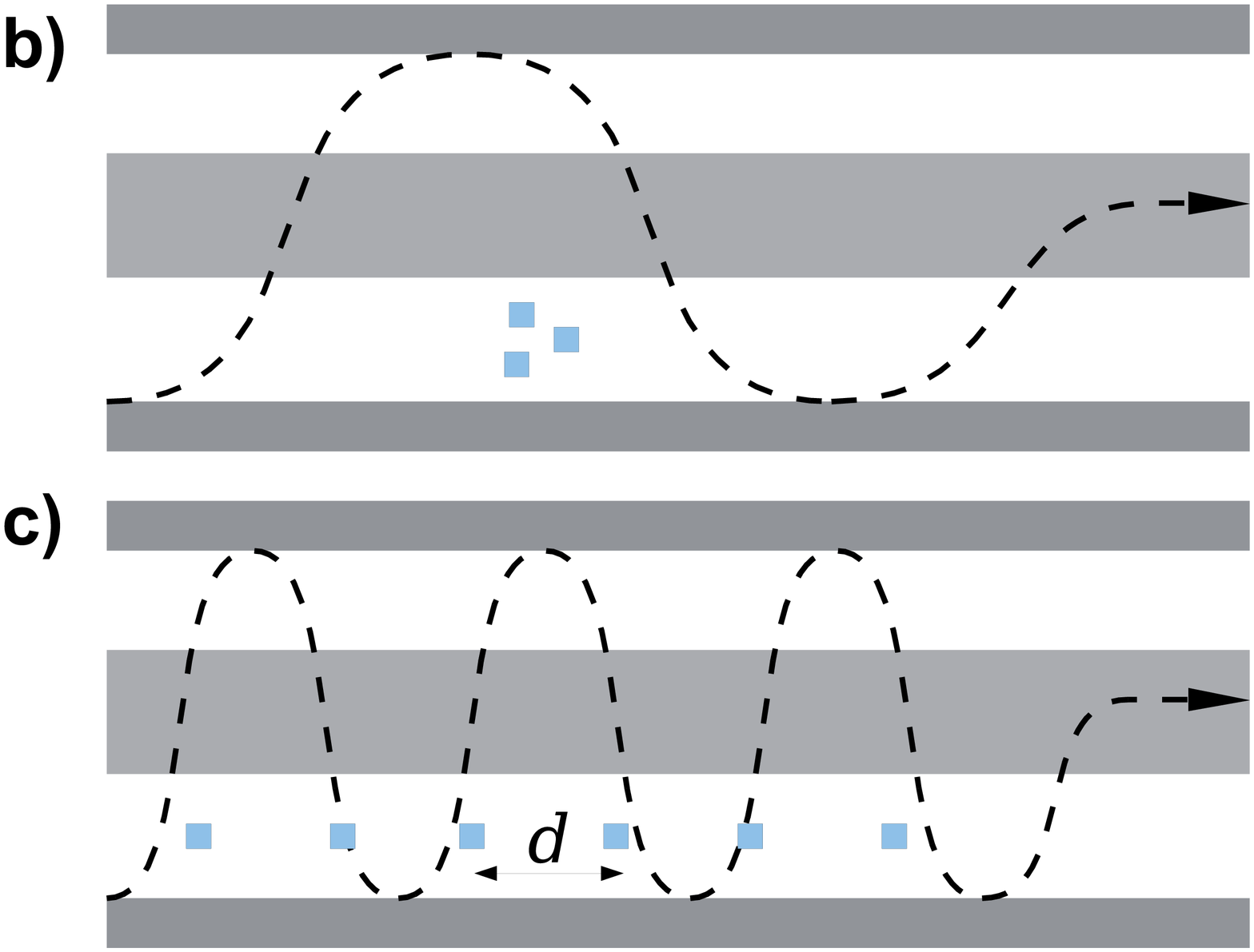}
\caption{Our photodetector scheme consists of three-level absorbers (a) embedded in a coplanar coaxial microwave guide (long gray stripes). The absorbers (squares) can either be grouped in clusters that are much smaller than a wavelength (b) or
regularly spaced (c). As the waveguide provides an effectively one-dimensional setup, the transverse position of the absorbers affects mildly the coupling strength.}
\label{fig:schema}
\end{figure}

Our proposal for a photon detector consists of a very simple setup, a
microwave guide plus a number of superconducting circuits that absorb photons
[Fig.~\ref{fig:schema}], and is able to circumvent at once the problems describe above. Instead of unitary evolutions, we make use of an irreversible
process which maps an excitation of the travelling electromagnetic field (a
photon) into an excitation of a localized quantum circuit. By
separating this encoding process from the later readout of this information, we avoid the backaction problem coming from continuous measurement. Furthermore, we can compensate different limitations ---weak coupling, low efficiency of absorption, photon bandwidth--- using no more than a few absorbing elements~[Fig.~\ref{fig:schema}c], where collective effects enhance the detection efficiency.

In this work we develop in great detail the theory underlying our proposal for high-efficiency phodetection~\cite{romero09}. In Sec.~\ref{sec:model}, we develop an abstract model that consists on a one-dimensional wave guide that transports photons and a number of three-level quantum systems that may absorb those photons. We will solve analytically the evolution of an incoming wavepacket, studying the time evolution of the full system with one or more detecting elements. For a single photon we can compute the absorption efficiency and how it depends on various parameters: detuning of photons from the absorber transition, irreversible decay rate of absorbers, spatial distribution of detectors and temporal profile of photon. In Sec.~\ref{sec:application}, we go further with this theoretical model by applying it to a particular design where the waveguide is a coplanar coaxial microwave guide, built with superconducting materials, and the absorbing elements are built from superconducting phase qubits. The parameters of the abstract model are matched to the experimental frequencies, impedances and capacitances of the chosen setup, verifying the experimental viability of our model. For completeness, in Sec.~\ref{sec:waveguide}, we write down
the quantum field theory for the microwave guide as it is used and needed in Sec.~\ref{sec:model} and Sec.~\ref{sec:application}. Finally, in Sec.~\ref{sec:final}, we summarize our results and discuss the experimental challenges and future scopes.

\section{Abstract model and design}
\label{sec:model}

In this section we present in an abstract manner our photodetector design: a one-dimensional waveguide transporting photons that can be absorbed by one or more bistable elements that, for the sake of simplicity, will be called indistinctively absorbers or qubits. We will introduce a general mathematical model that uses
two fields to describe the quantized waves (photons) and three-level
systems for the absorbers. We will study the consequences of this
model, including how the joint device may act as a photon detector or
photon counting device. This mathematical formalism will serve as a
foundation for the developments in Sec.~\ref{sec:application}, where we will specify the
microscopic theory behind the waveguide and the photon absorbers.

\subsection{Master equation}

Following the standard quantum optical master equation formalism, we describe the joint state of the propagating photons and the absorbing elements using a density matrix, $\rho.$ This is a Hermitian operator that contains information about the probability distributions of all observables in the system. To the lowest order of approximation the density matrix will evolve according to the master equation
\begin{equation}
\label{master-eq}
\frac{d}{dt}\rho = -\frac{i}{\hbar}[H, \rho] + {\cal L}\rho .
\end{equation}
Here, the conservative part of the
evolution is ruled by the Hamiltonian
\begin{equation}
 H = H_{\rm photon} + H_{\rm qubit} + H_{\rm int},
\end{equation}
which contains terms describing the propagation of microwave photons
in the guide, the absorbing elements and the interaction between
matter (absorbers) and radiation (field), respectively.  The first operator contains
two propagating fields, $\psi_r$ and $\psi_l$ moving rightwards and
leftwards with group velocity $v$
\begin{equation}
  \label{model-photon}
  H_{\rm photon} =\int \left[\psi_r^\dagger(-i\hbar v\partial_x)\psi_r +
    \psi_l^\dagger(+i\hbar v\partial_x)\psi_l\right]dx.
\end{equation}
The second part models one or more discrete quantum elements that we
place close to the transmission line. These elements are analogous to the
qubits in quantum computing and circuit-QED setups, and will play the role of absorbers, or qubits, enjoying at least three energy levels. The
first two, $\ket{0}$ and $\ket{1},$ are metastable and separated by an
energy $\hbar\omega$ close to the frequency of the incoming photons
\begin{eqnarray}
  \label{model-qubit}
  H_{\rm qubit} = \sum_{i=1}^N \hbar\omega \ket{1}_i\bra{1}.
\end{eqnarray}
Then, there is the interaction between the electromagnetic field and
our qubits. We model it with a delta-potential which induces
transitions between the qubit states at the same time it steals or
deposits photons in the wave guide
\begin{equation}
  \label{model-interaction}
  H_{\rm int}=\sum_{i=1}^N \int V\delta(x-x_i) [\psi_r(x) + \psi_l(x)]\ket{1}_i\bra{0}dx
  + \mathrm{H.c.}
\end{equation}
Finally we have included a Liouvillian operator ${\cal L}$ which
models the decay of the absorbing elements from the metastable state
$\ket{1}$ to a third state, $\ket{g},$ and which constitutes the
detection process itself. A general second order Markovian model
for the decay operator reads
\begin{equation}
  {\cal L}\rho = \sum_{i=1}^N  \frac{\Gamma}{2}\left[
    2\ket{g}_i\bra{1}\rho\ket{1}_i\bra{g} -
    \ket{1}_i\bra{1}\rho - \rho\ket{1}_i\bra{1}\right].
\end{equation}
Note that if we start with a decoupled qubit $(V=0)$ in state
$\ket{1}$, the population of this state is depleted at a rate
$\Gamma$
\begin{equation}
  \rho(t) = e^{-\Gamma t}\ket{1}\bra{1} + \ldots.
\end{equation}

\subsection{Non-Hermitian solution}

Let us consider the simple case of one qubit or absorber. The master
equation (\ref{master-eq}) can be written in a more convenient form
\begin{equation}
  \frac{d}{dt}\rho = A \rho + \rho A^\dagger + \Gamma \ket{g}\bra{1}\rho
  \ket{1}\bra{g},
\end{equation}
where we have introduced a non-Hermitian operator
\begin{eqnarray}
  A &=& -\frac{i}{\hbar} H - \frac{\Gamma}{2}\ket{1}\bra{1} \neq A^\dagger.
\end{eqnarray}
The master equation can now be manipulated formally using the
``interaction'' picture
\begin{equation}
  \rho(t) = e^{At} \sigma(t) e^{A^\dagger t},
\end{equation}
with the following equation for $\sigma(t),$
\begin{equation}
  \frac{d}{dt}\sigma = \Gamma e^{-A t}\ket{g}\bra{1}e^{At}
  \sigma e^{A^\dagger t}\ket{1}\bra{g} e^{-A^\dagger t}.
\end{equation}
Using the relation $e^{At}\ket{g} = e^{A^\dagger t}\ket{g} = \ket{g}$
we obtain
\begin{equation}
  \frac{d}{dt}\sigma = \Gamma \ket{g}\bra{1}e^{At}
  \sigma e^{A^\dagger t}\ket{1}\bra{g}.
\end{equation}
This equation can be solved by splitting into a part that already
decayed, $\sigma_{gg}\propto \ket{g}\bra{g},$ and all other
contributions $\sigma_{\perp} = 1 - \sigma_{gg}$
\begin{eqnarray}
  \frac{d}{dt}\sigma_{gg} &=& \Gamma \ket{g}\bra{g}
  \times \sum_{\alpha,\beta} \bra{1}e^{At}\sigma_{\perp}e^{A^\dagger t}\ket{1},\\
  \frac{d}{dt}\sigma_{\perp} &=& 0.
\end{eqnarray}
Using the initial condition
\begin{equation}
  \sigma_{gg}(0) = 0,\; \sigma_{\perp}(0) = \rho(0) = \ket{\Psi(0)}\bra{\Psi(0)},
\end{equation}
we obtain the formal solution
\begin{equation}
  \rho(t) = \ket{\Psi(t)}\bra{\Psi(t)} + \ket{g}\bra{g}
  \int_{0}^{t} \Gamma \braket{1\vert\Psi(\tau)}\braket{\Psi(\tau)\vert1}d\tau,
  \label{dty-matrix-sol}
\end{equation}
where we have introduced state $\ket{\Psi(t)}$, evolving from initial state $\ket{\Psi(0)}$ according to the lossy Schr\"odinger equation
\begin{equation}
\label{eq:lossy}
\frac{d}{dt}\ket{\Psi(t)} = A \ket{\Psi(t)} \equiv -\frac{i}{\hbar}\bar{H}
\ket{\Psi(t)} .
\end{equation}
Note that the probability of the qubit irreversibly capturing a photon is given by
\begin{eqnarray}
  P(0\to1\to g) &=& \mathrm{Tr}(\rho\ket{g}\bra{g}) \\
  &=& 1 - \mathrm{Tr}\left[\ket{\Psi(t)}\bra{\Psi(t)}\right] \nonumber\\
  &=& 1 - \Vert\Psi\Vert^2.\label{norm-loss}
\end{eqnarray}

\subsection{Model equations for a single absorber}

Following the previous reasoning, our goal is to solve the lossy
Schr\"odinger equation
\begin{equation}
\label{non-Hermitian}
i \frac{d}{dt}\ket{\Psi} = \bar H \ket{\Psi},
\end{equation}
ruled by the non-Hermitian operator
\begin{equation}
\bar H = H_{photon} + H_{qubit} + H_{int} -\sum_i \frac{\hbar\Gamma}2\ket{1}_i\bra{1}.
\end{equation}
In order to describe the detection process, our initial condition
$\ket{\Psi(0)}$ will contain the incoming photons plus all absorbers
in the ground state $\ket{0}$. The probability that an absorber
captures a photon and changes from state $\ket{0}$ to $\ket{g}$ is
given by the loss of norm shown in Eq.~(\ref{norm-loss}).

Let us consider an incident photon coming from the left with energy
$E=\hbar|k|v$. The state of the system will be given by~\cite{shen05}
\begin{eqnarray}
\ket{\Psi} &=& \int \left[ \xi_r(x,t) \psi_r^\dagger(x) +
\xi_l(x,t) \psi_l^\dagger(x)\right] \ket{vac,0} +\nonumber\\
& + & e_1 \ket{vac,1} .
\end{eqnarray}
Here $\psi_{r,l}^\dagger(x)\ket{vac,0}$ is the state of a photon
created at position $x$ and moving either to the right or to the left,
while the absorber is in metastable state $\ket{0}$. Also, $\xi_r(x,t)$ and
$\xi_l(x,t)$ represent the wavefunction of a single photon moving to
the right and to the left, respectively, while $\ket{vac,1}$ is a state with no
photons and the absorber excited to unstable level $\ket{1}.$ Note
that thanks to the relation~(\ref{dty-matrix-sol}) we do not need to
explicitely include the population of state $\ket{g}.$ We only have to
solve the Schr\"odinger equation (\ref{non-Hermitian}) using a
boundary condition that represents a photon coming from the left and
an inactive absorber
\begin{equation}
  \xi_r(x>0,t_0) = \xi_l(x,t_0) = 0,\; e_1(t_0) = 0,
\end{equation}
and compute the evolution of the photon amplitude, $\xi_{r,l}(x,t),$ the excited
state population $e_1(t)$ and the resulting photon absorption probability (\ref{norm-loss}).

\begin{figure}[t]
\centering
\includegraphics[width=0.5\linewidth]{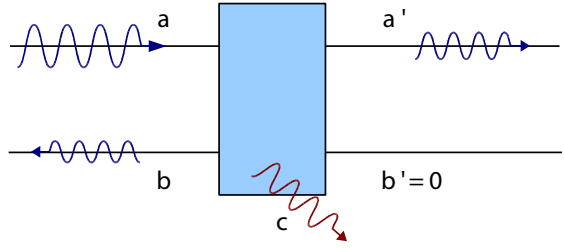}
\caption{An incident photon, moving rightwards, interacts with an absorbing element. Out of the original amplitude of the field, $a,$ a component is transmitted, $a',$ another component is reflected, $b,$ and finally with some probability, $|c|^2,$ the system absorbs a photon and changes state.}
\label{fig:scattering1}
\end{figure}

After decomposing the wave equation into the left $(x < 0)$ and right
$(x > 0)$ halves of space, and replacing the potential $\delta(x-x_1)$
with an appropriate boundary condition at $x_1=0,$ one obtains the
Schr\"odinger equation for the absorber
\begin{eqnarray}
  i\partial_t e_1 &=&\frac{V}{2\hbar}\left[
    \xi_r(0^+) + \xi_r(0^-) + \xi_l(0^+) + \xi_l(0^-)\right],\nonumber\\
  &+& (\omega-i\Gamma)e_1
\end{eqnarray}
and four equations for the photon,
\begin{eqnarray}
  i\partial_t \xi_r(x,t) &=& -iv\partial_x\xi_r(x,t),\quad x\neq 0 , \\
  i\partial_t \xi_l(x,t) &=& +iv\partial_x\xi_l(x,t),\quad x\neq 0 , \nonumber\\
  0 &=& -i\hbar v[\xi_r(0^+,t) - \xi_r(0^-,t)] + Ve_1(t) , \nonumber\\
  0 &=& +i\hbar v[\xi_l(0^+,t) - \xi_l(0^-,t)] + Ve_1(t) . \nonumber
\end{eqnarray}
We introduce new variables $a, b, a'$ and $b'$ describing the
amplitude of the fields on both sides of an absorber
[Fig.~\ref{fig:scattering1}] ,
\begin{equation}
  \begin{array}{ll}
  a(t) = \xi_r(0^-,t),& a'(t) = \xi_r(0^+,t),\\
  b(t) = \xi_l(0^-,t),& b'(t) = \xi_l(0^+,t).
  \end{array}
  \label{amplitudes}
\end{equation}
Two of these variables can be solved from the initial conditions
\begin{eqnarray}
  a(t) &=& \xi_r(0^+,t) = \xi_r(-v(t-t_0),t_0)\,\label{incoming}\\
  b'(t) &=& 0,\nonumber
\end{eqnarray}
while the rest are extracted from the boundary conditions
\begin{equation}
  a'(t) = a(t) + \frac{V}{i\hbar v} e_1(t),\;
  b(t) = \frac{V}{i\hbar v} e_1(t).\label{boundary}
\end{equation}
We have expressed all unknowns in terms of $e_1(t).$ The population of
the excited state now satisfies
\begin{eqnarray}
i \partial_t e_1 &=&
\left(\omega - i\Gamma - i\frac{V^2}{\hbar^2v}\right) e_1
+ \frac{V}{\hbar} a(t),
\end{eqnarray}
which contains the original frequency, $\omega,$ and an imaginary
component for the decay of the three-level system, $\Gamma,$ enhanced
by the interaction with the transmission line $V^2/\hbar^2v.$ Solving for $e_1$, we find
\begin{eqnarray}
  e_1(t) &=& -i\frac{V}{\hbar} \int_0^t e^{-\tilde\Gamma (t-\tau)}a(\tau)d\tau, \\
 \tilde\Gamma &=& \Gamma + \frac{V^2}{\hbar^2 v} + i \omega,\nonumber
\end{eqnarray}
in terms of the incoming beam (\ref{incoming}).

Using the expression of $e_1(t)$ one may compute the reflected and
transmitted components of a microwave beam, to find that the total
intensity is not conserved. Rather, we have an absorbed component that
is ``stolen'' by the three-level system to perform a transition from
the $\ket{0}$ to $\ket{g}$ states via $\ket{1}.$ The balance equations
are thus [Fig.~\ref{fig:scattering1}]
\begin{equation}
  |a|^2 + |b'|^2 = |a'|^2 + |b|^2 + |c|^2.
\end{equation}
If we consider a photon with a finite duration, $T,$ the detection
efficiency will be defined as the fraction of the wavepacket that was
absorbed
\begin{equation}
  \alpha = \frac{\int_0^T |c(t)|^2 dt}
  {\int_0^T\left[|a(t)|^2 + |b'(t)|^2\right]dt} \, .
\end{equation}
For the purposes of photodetection and photon counting we would like
this value to reach the maximum $\alpha = 1.$

\subsection{Stationary solutions for one qubit}

To understand the photon absorption process we will study a continuous
monochromatic beam which is slowly switched on,
\begin{equation}
  a(t) = \exp[(-i\omega_0 + \sigma) t]\quad t \leq 0,\sigma>0.
\end{equation}
Taking the limits $t_0 \to -\infty$ and $\sigma\to0^+$ in this precise
order, we obtain
\begin{eqnarray}
  e_1 &=& -i\frac{V}{\hbar}\int_{-\infty}^t
  \exp[-\tilde \Gamma (t - \tau) + (-i\omega_0 + \sigma)\tau]\nonumber\\
  &=& -i \frac{V/\hbar}{\Gamma + \frac{V^2}{\hbar^2 v}
    + i (\omega - \omega_0)} e^{-i\omega_0 t}.
\end{eqnarray}
We will introduce a unique parameter to describe the photodetection process,
\begin{equation}
  \gamma = \frac{\hbar v}{V^2} \hbar\left[\Gamma + i (\omega - \omega_0)\right]
  \label{renormalized-gamma} ,
\end{equation}
which includes both a renormalization of the decay rate and a small imaginary component associated to the detuning. With this parameter the solution reads
\begin{eqnarray}
  e_1(t) &=& -i \frac{\hbar v}{V} \frac{1}{1 + \gamma} e^{-i\omega_0 t}\\
  a'(t) &=& \left[1 -  \frac{1}{1+\gamma}\right] e^{-i\omega_0 t},\\
  b(t) &=& -\frac{1}{1 + \gamma}e^{-i\omega_0 t}.
\end{eqnarray}
If we work in the perfectly tuned regime, $\omega=\omega_0,$ the decay
rate $\gamma$ becomes real and the absorption rate is
\begin{eqnarray}
  \alpha &=& - 2 \mathrm{Re}(a^* b) - 2|b|^2 \\
  &=& \frac{2}{1+\gamma} - \frac{2}{(1+\gamma)^2}\\
  &=&\frac{2\gamma}{(1+\gamma)^2} = 2b(1 - b).
\end{eqnarray}
This value achieves a maximum of $50\%$ efficiency or $\alpha=1/2$ at
the values $b=1/2,$ $\gamma=1$. We think that the limit of $50\%$ in the photodetection efficiency is fundamental and related to the Zeno effect, expressing the balance of quantum information between, see Fig.~1a, the reversible absorption of the photon in the first (left) transition channel and the irreversible absorption in the second (rigth) one. 

\subsection{Transfer matrix}

We can derive the long wavepacket or quasi-stationary solution in a
slightly different manner. Note that for infinitely long wavepackets
the population of the excited state is determined by the fields on
both sides
\begin{eqnarray}
  e_1 &=& \frac{1}{(\omega_0 - \omega)+i\Gamma}\frac{V}{2\hbar}
  \left[a + a' + b + b'\right] \\
  &=& \frac{\hbar v}{iV\gamma}\left[a + a' + b + b'\right].\nonumber
\end{eqnarray}
With this the boundary conditions in Eq.~(\ref{boundary}) transform
into a set of equations that only involves the incoming and outgoing
fields,
\begin{eqnarray}
  0 = a' - a + \frac{1}{2\gamma}(a + a' + b + b'),\\
  0 = b' - b - \frac{1}{2\gamma}(a + a' + b + b').
\end{eqnarray}
In terms of the matrix and vectors
\begin{equation}
  A = \left(\begin{array}{cc} 1 & 1 \\ -1 & -1\end{array}\right),\;
  x = \left(\begin{array}{c} a \\ b\end{array}\right),\;
  x'= \left(\begin{array}{c} a' \\ b'\end{array}\right),
\end{equation}
we can write
\begin{equation}
  \left(1 + \frac{1}{2\gamma}A\right) x' = \left(1 - \frac{1}{2\gamma}A\right)x.
\end{equation}
Multiplying everything by $(1 - A/2\gamma)$ and using $A^2=0$ gives
$x' = \left(1 - A/\gamma\right)x.$ This amounts to a relation between
the fields to the right and to the left of the absorber
\begin{equation}
  \left(
    \begin{array}{c}
      a' \\ b'
    \end{array}
    \right) =
    T
  \left(
    \begin{array}{c}
      a \\ b
    \end{array}
    \right),
  \label{transfer-matrix}
\end{equation}
given by the scattering matrix
\begin{equation}
  T =   \left(
    \begin{array}{cc}
      1 - \frac{1}{\gamma} & - \frac{1}{\gamma} \\
      \frac{1}{\gamma} & 1 + \frac{1}{\gamma}
    \end{array}
    \right) = 1 - \frac{1}{\gamma}A.
\end{equation}
Note that the value $\gamma$ introduced
before~(\ref{renormalized-gamma}) only depends on the properties of
the qubit, the interaction between the qubit and the line and the
group velocity. We have thus elaborated a rather compact and easily
generalizable scheme for studying the scattering and absorption of
photons by the individual absorbers.

\subsection{General absorption formula}

Using the transfer matrix formalism we can compute the fraction of
absorbed photons. The only requirement to obtain a general formula is
that the scattering matrix remains the same under inversion
\begin{equation}
  \left(
    \begin{array}{c}
      b \\ a
    \end{array}
    \right) =
    T
  \left(
    \begin{array}{c}
      b' \\ a'
    \end{array}
    \right).
\end{equation}
This is reasonable: from Fig.~\ref{fig:scattering1} we can conclude
that $b$ and $a'$ have the same role, just like $b'$ and $a.$ Using
the relations $b = T_{01} a'$ and $a = T_{11} a',$ the absorbed
fraction becomes
\begin{equation}
  \alpha = 1 - \frac{|a'|^2 + |b|^2}{|a|^2} = 1 - \frac{1}{|T_{11}|^2}
  (1 + |T_{01}|^2).
  \label{general-alpha}
\end{equation}
This is consistent with the single absorber case
\begin{eqnarray}
  T_{11} = \frac{1+\gamma}{\gamma},\; T_{01} = -\frac{1}{\gamma},
\end{eqnarray}
in which we recover
\begin{equation}
  \alpha = 1 - \frac{\gamma^2}{(1+\gamma)^2}\left(1+\frac{1}{\gamma^2}\right)
  =\frac{2\gamma}{(1+\gamma)^2},
  \label{abs-1q}
\end{equation}
as expected.

\begin{figure}[t]
\centering
\includegraphics[width=0.5\linewidth]{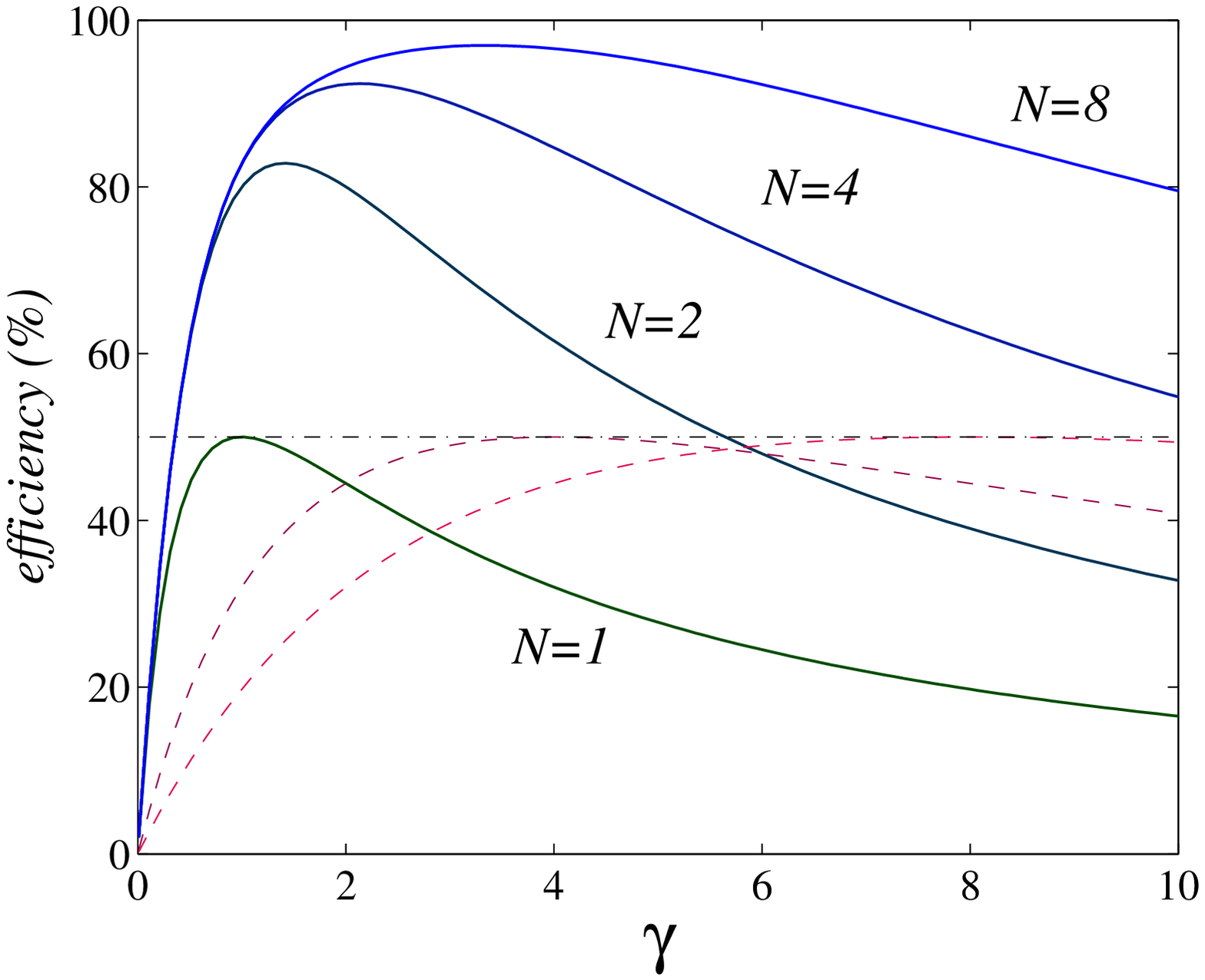}
\includegraphics[width=0.5\linewidth]{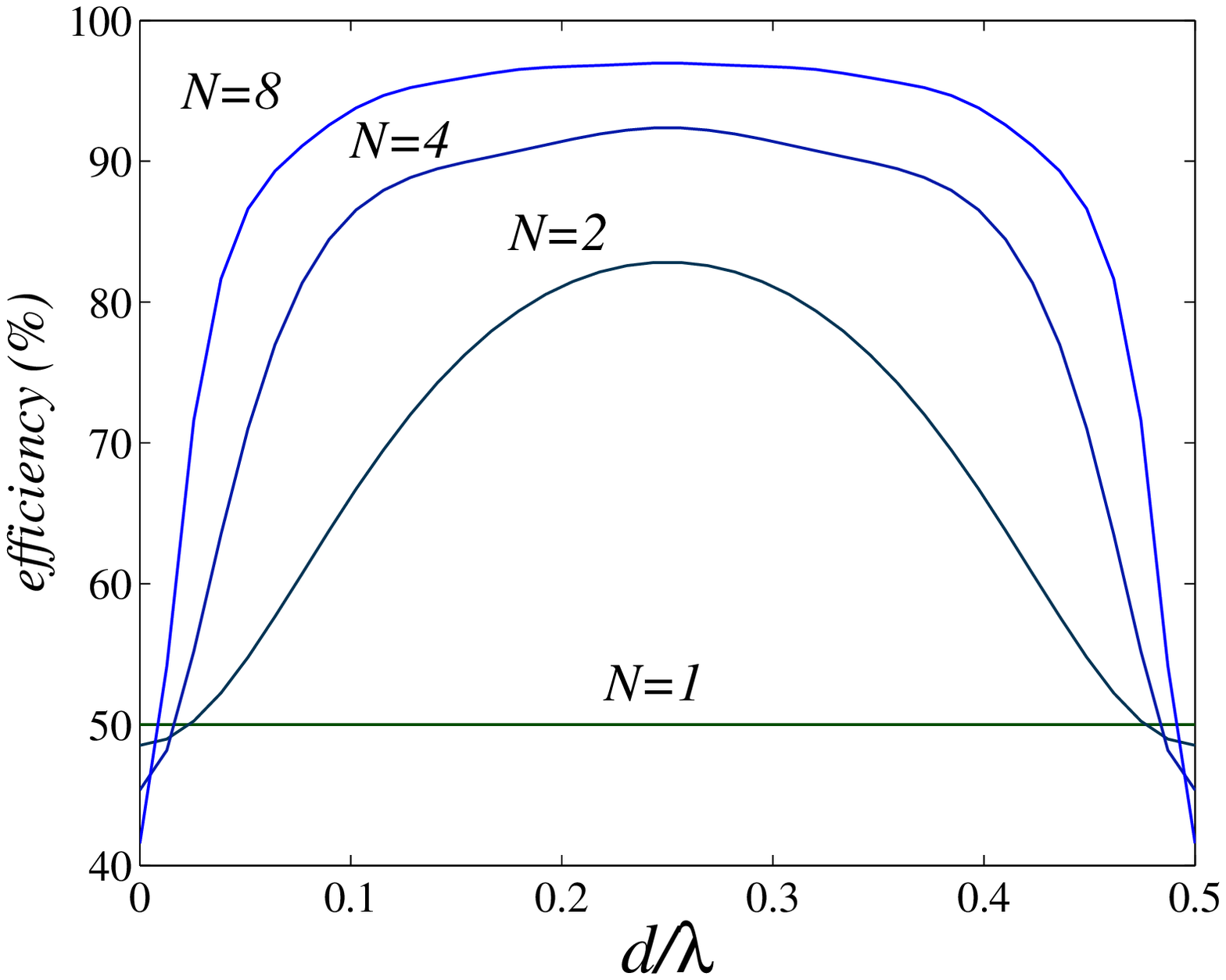}
\caption{Detection efficiency. (Top) Absorption probability
  vs. effective decay rate for a setup with $N=1, 2, 4$ and $8$ qubits
  (black, green, blue, red) either in cluster (dashed) or array
  (solid) regime. In the first case, where qubits are close together,
  the efficiency is fundamentally limited to $50\%,$ while in the
  other case there is no upper limit. In both cases, having more
  qubits allow us to use weaker couplings or stronger decay
  rates. (Bottom) Detection efficiency versus the separation $d$ of
  the absorbing elements for an array of periodically distributed
  Josephson junctions. All curves have been produced for resonant
  qubits.}
\label{fig:efficiencies}
\end{figure}

\subsection{Multiple absorbers}

If we have more than one absorbing element at positions $x_1, x_2
\ldots$, we will use the same formula for the absorption
efficiency~(\ref{general-alpha}), but with transmission matrix
\begin{equation}
  T_{total} = T_1 \Phi(x_2-x_1) T_2 \Phi(x_3-x_2) \cdots T_N.
\end{equation}
Here, $T_i$ describes the absorption properties of a given absorber
and thus depend on the value of $\gamma_i$, while the phase matrix
$\Phi(x)$ is given by
\begin{equation}
  \Phi(x) = \left(
    \begin{array}{cc}
      e^{i\theta(x)} & 0\\
      0 & e^{-i\theta(x)}
    \end{array}\right)
\end{equation}
and contains the accumulated phase of the electromagnetic wave when
travelling between consecutive absorbers
\begin{equation}
    \theta(x) = \frac{2\pi}{\lambda} x = \frac{\omega_0}{v} x \, .
\end{equation}

\begin{figure}[t]
\centering
\includegraphics[width=0.7\linewidth]{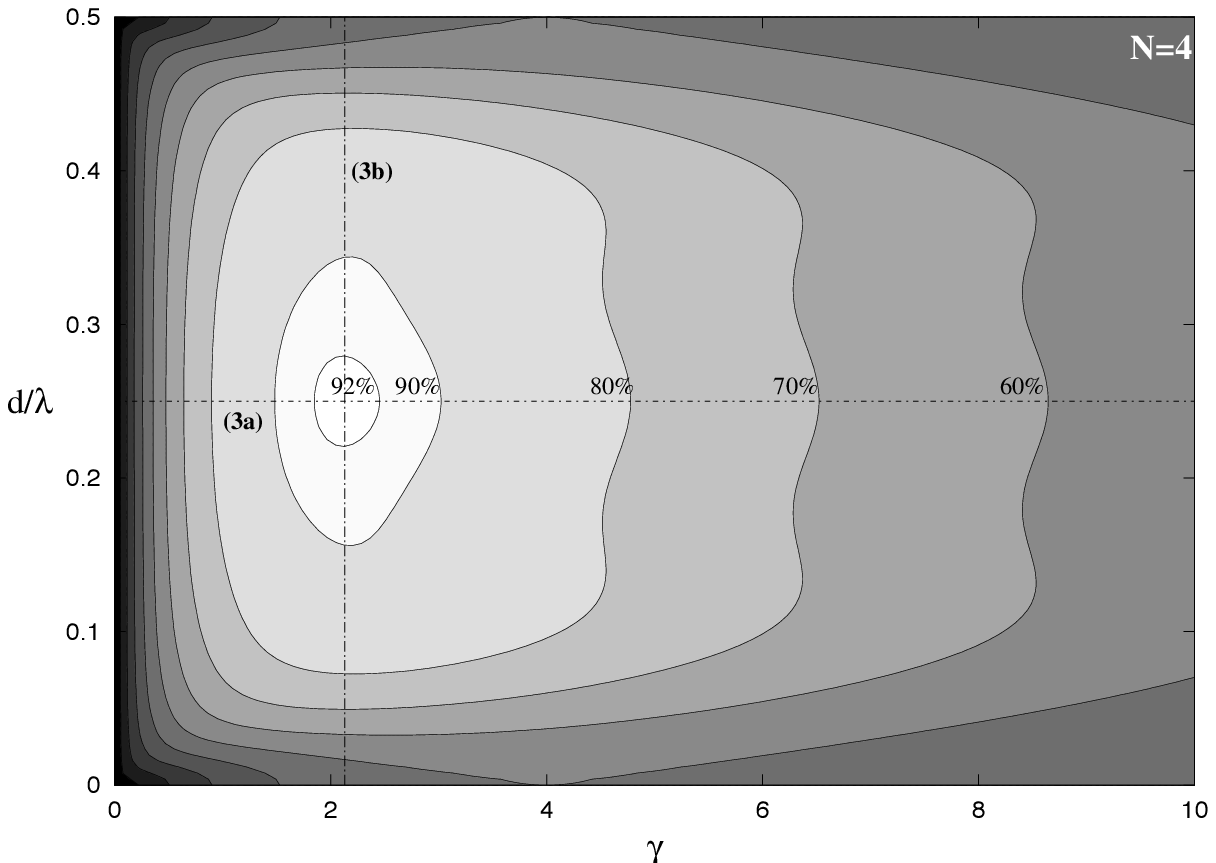}
\includegraphics[width=0.7\linewidth]{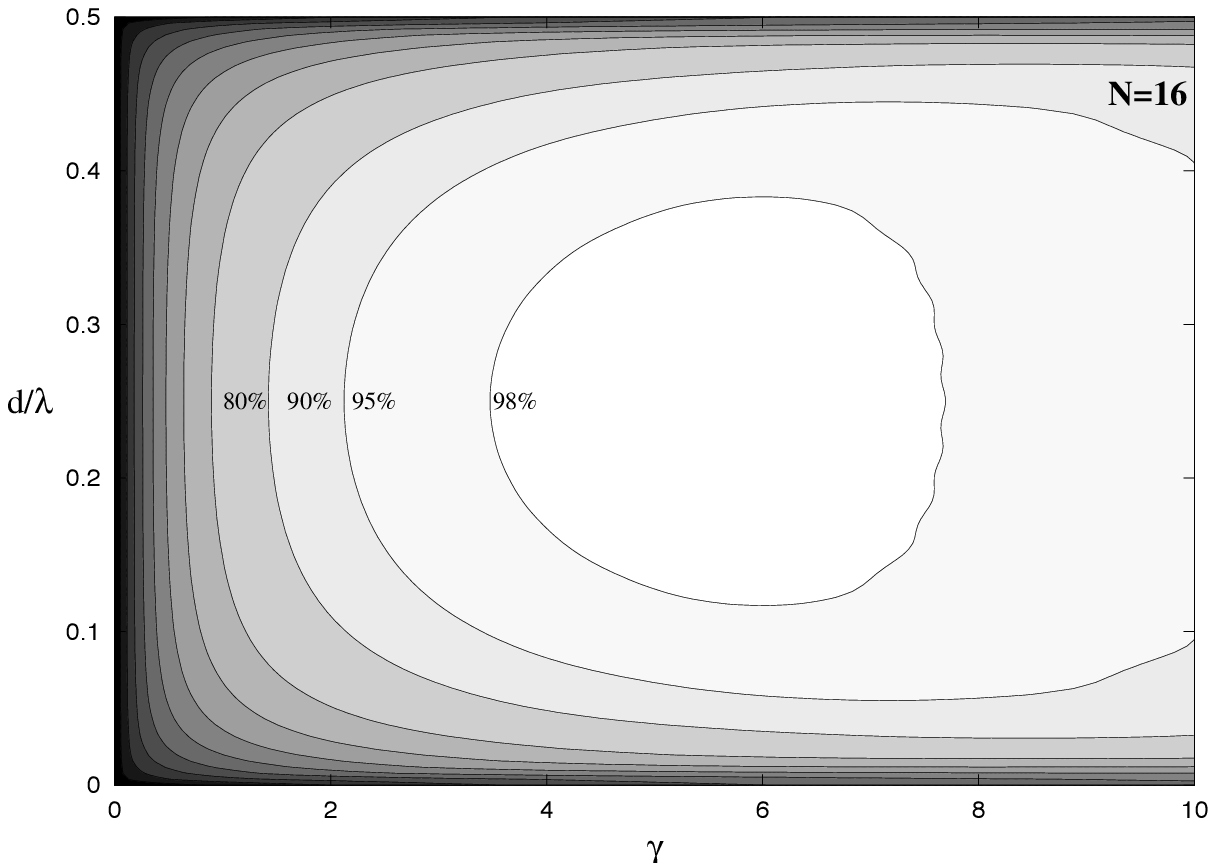}
\caption{Dependency of the detector efficiency on the parameter
  $\gamma$ and on the separation between absorbers, in units of the
  microwave wavelength, $d/\lambda.$ We have chosen $N=4$ (above) and
  $N=16$ (below). Note that the maximum gets broader with $N,$ but it
  is always achieved at $d=\lambda/4.$ The solid lines denote the cuts
  corresponding to the plots in Fig.~\ref{fig:efficiencies}.}
\label{fig:contour}
\end{figure}

\subsection{Analysis of the efficiency}

The relation (\ref{transfer-matrix}) is valid for one, two and more
absorbing elements, placed at different positions. If we have a single
element, we have already found the absorbed fraction
(\ref{abs-1q}). As shown in Fig.~\ref{fig:efficiencies}, this quantity
is upper bounded by $50\%$ which is reached for resonant qubits with
$\gamma=1.$ This means that a single absorbing element, without
knowing the arrival time and no particular external modulation, can
only detect in average half of the incoming photons.

The situation does not improve if we place $N$ absorbing elements
close together [Fig.~\ref{fig:schema}b]. When the wavelength of the
microwave field is large compared to the size of the absorber
cluster, the total transfer matrix is given by
\begin{equation}
  T = T_{N}^N = \left(1 + \frac{1}{\gamma}A\right)^N = 1 + \frac{N}{\gamma} A \, .
\end{equation}
This is just the same matrix with a smaller effective decay rate
$\gamma/N,$ and the total efficiency remains limited to $50\%$,
\begin{equation}
  \alpha_N = \frac{2(\gamma/N)}{[1 + \gamma/N]^2} \leq \frac{1}{2} \, .
\end{equation}

We can do much better, though, if we place a few absorbing elements
separated by some distance $d$ [Fig.~\ref{fig:schema}c]. In this case
the total transfer matrix is given by
\begin{equation}
T_{N}(d) = \left[T_1 \left(\begin{array}{cc}
e^{i\omega_0d/v} & 0 \\ 0 & e^{-i\omega_0d/v} \end{array}\right)\right]^N.
\end{equation}
The total efficiency now depends on two variables, $\gamma$ and the
phase $\theta=\omega_0d/v$ or the separation between absorbers, $d.$
As Fig.~\ref{fig:contour} shows, the optimal value of the phase is
$\theta=\pi/2,$ corresponding to $\lambda/4$ separation and, in this case,
the total efficiency is no longer limited. For instance, as shown in
Fig.~\ref{fig:efficiencies} for two and three qubits on-resonance the
efficiency can reach $80\%$ and $90\%,$ respectively. Not only it
grows, but it does so pretty fast.

\begin{figure}[t]
\centering
\includegraphics[width=0.7\linewidth]{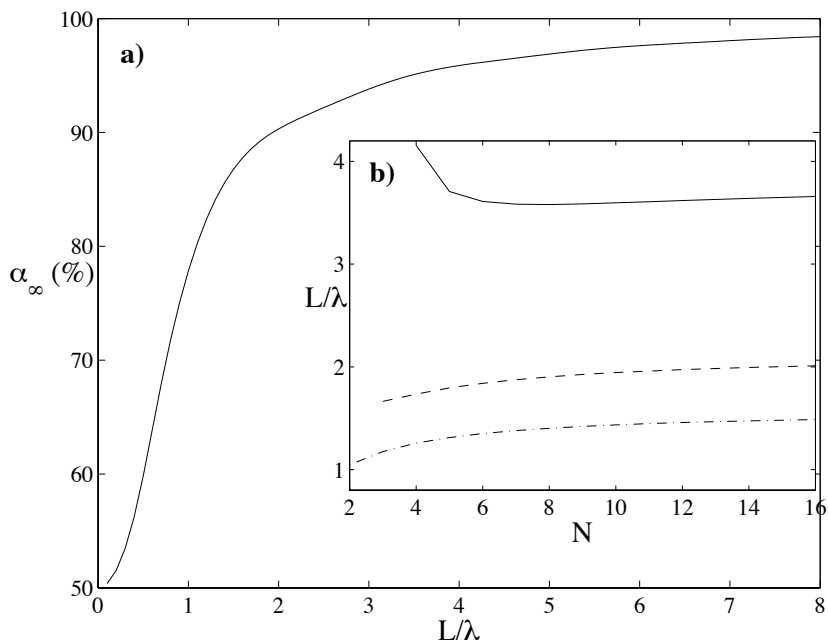}
\caption{Optimal separation of detectors.  (a) Asymptotic detector
  efficiency $\alpha_\infty(L)$ as a function of the setup size $L,$
  and computed as in Eq.~(\ref{asym-alpha}). (b) Minimal size $L$
  of a setup with $N$ qubits required to attain a detection efficiency
  $\alpha=78\%$ (dash-dot), $85\%$ (dashed) or $\alpha=92\%$ (solid).}
\label{fig:separation}
\end{figure}

An interesting feature is that, as we increase the number of qubits,
the absorbed fraction becomes less sensitive to the qubit
separation, which allows for more compact setups than one would
otherwise expect. For instance, in Fig.~\ref{fig:separation} we show
the total size of a setup computed for a fixed detection efficiency
and a given number of qubits. Since the optimal separation behaves as
$d \sim 1/N,$ the total system size for a fixed detector efficiency,
$78\%, 80\%$ and so on, remains bounded, even for large number of
qubits. Moreover, as one increases the number of qubits, the
efficiency grows rapidly. We can define the value
\begin{equation}
  \label{asym-alpha}
  \alpha_\infty({\cal L}) =
  \lim_{N\to\infty}\max_\gamma\alpha({\cal L},N,\gamma,d={\cal L}/N) ,
\end{equation}
which gives an idea of what is the maximum detection efficiency for a
given circuit size. The value shown in Fig.~\ref{fig:separation}b
approaches the limit of 100\% quite fast and gives us an idea of the
\textit{minimal} size of a detector which is needed to obtain a given
efficiency.

\subsection{Robustness against imperfections}

There are many factors that will condition the actual efficiency of a
photodetector. Some of them will have to be discussed later on in the
context of the proposed implementation, but others can be analyzed
already with the present theory.

The first source of errors that one may consider are systematic
differences in the fabrication and tuning of the absorbing
elements. These fluctuations are currently unavoidable, and may even
evolve through the lifetime of a setup, due to changes in the
temperature, fluctuations of impurities, among others. These systematic errors
could be modeled by random perturbations in the parameters of the
three-level systems, either due to inhomogeneous broadening (different
frequencies $\omega_i$), inhomogeneous decay rates (different
$\Gamma_i$) or changes in the coupling strengths ($V_i$). However,
since the scattering of photons is described by a single parameter
$\gamma_i$ per qubit, it is more convenient to model the errors as
random changes in these values.

\begin{figure}[t]
\centering
\includegraphics[width=0.7\linewidth]{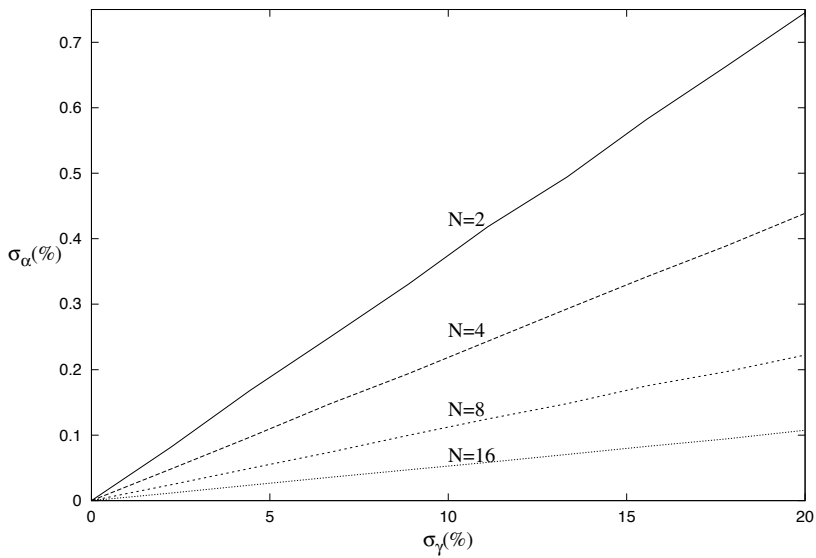}
\caption{Sensitivity of the detector to systematic errors in the
  absorbers. We plot the standard deviation of the absorption
  efficiency, $\sigma_\alpha,$ in percentiles, vs. the maximum
  systematic random error, $\sigma_\gamma,$ averaged over $10000$
  random samples with $N=4, 8, 16$ and 32 qubits (top to bottom). }
\label{fig:noise}
\end{figure}

We have performed a numerical study using a constant probability
distribution, $P(\gamma) = 1/2\sigma_\gamma\bar\gamma$ for $|\gamma -
\bar\gamma| \leq \sigma_\gamma\bar\gamma,$ with a maximum relative
error $\sigma_\gamma$ that ranges from 0 up to 20\%. Using 10000
random realizations with different parameters $\{\gamma_i\}_{i=1}^N,$
we made statistics of the detector efficiency, $\alpha.$ The results
are shown in Fig.~\ref{fig:noise}, where we plot the standard
deviation of the maximum detection efficiency, $\sigma_\alpha,$ given
by
\begin{equation}
  \sigma_\alpha = \sqrt{\langle\langle\alpha^2\rangle\rangle -
    \langle\langle\alpha\rangle\rangle^2},
\end{equation}
where the double bracket $\langle\langle\cdots\rangle\rangle$ denotes
average over the random instances. According to our results, the
maximum efficiencies for $N=4,8,16$, and $32$ are
$\alpha=92.4\%,97\%,98.9\%$ and $99\%,$ and in the worst case of
$20\%$ maximum systematic errors, where one standard deviation implies an
efficiency decrease of $0.7\%$ for $4$ absorbers and $0.1\%$ for
32. In other words, we find up to two orders of magnitude smaller errors in
the detector than in the fabrication.

\subsection{Detector bandwidth and dephasing}

\begin{figure}[t]
\centering
\includegraphics[width=0.7\linewidth]{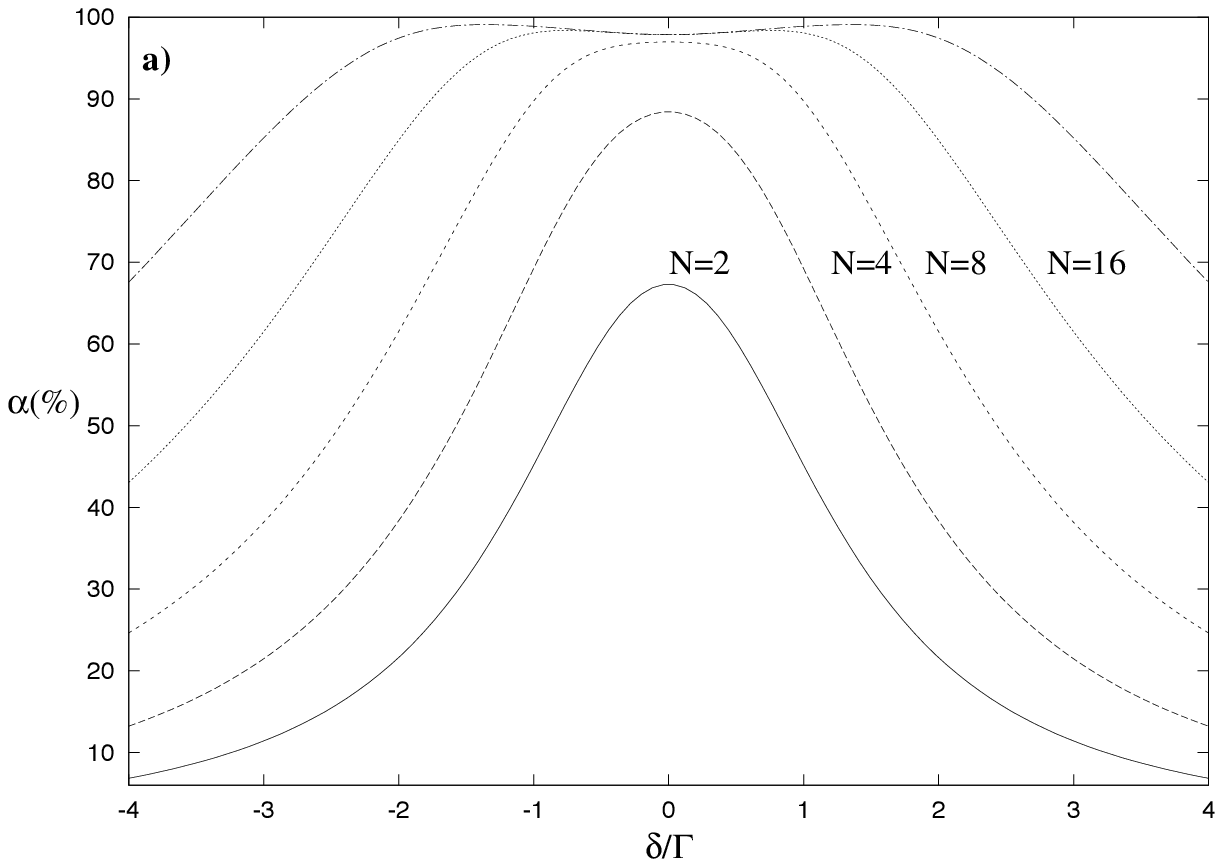}
\includegraphics[width=0.7\linewidth]{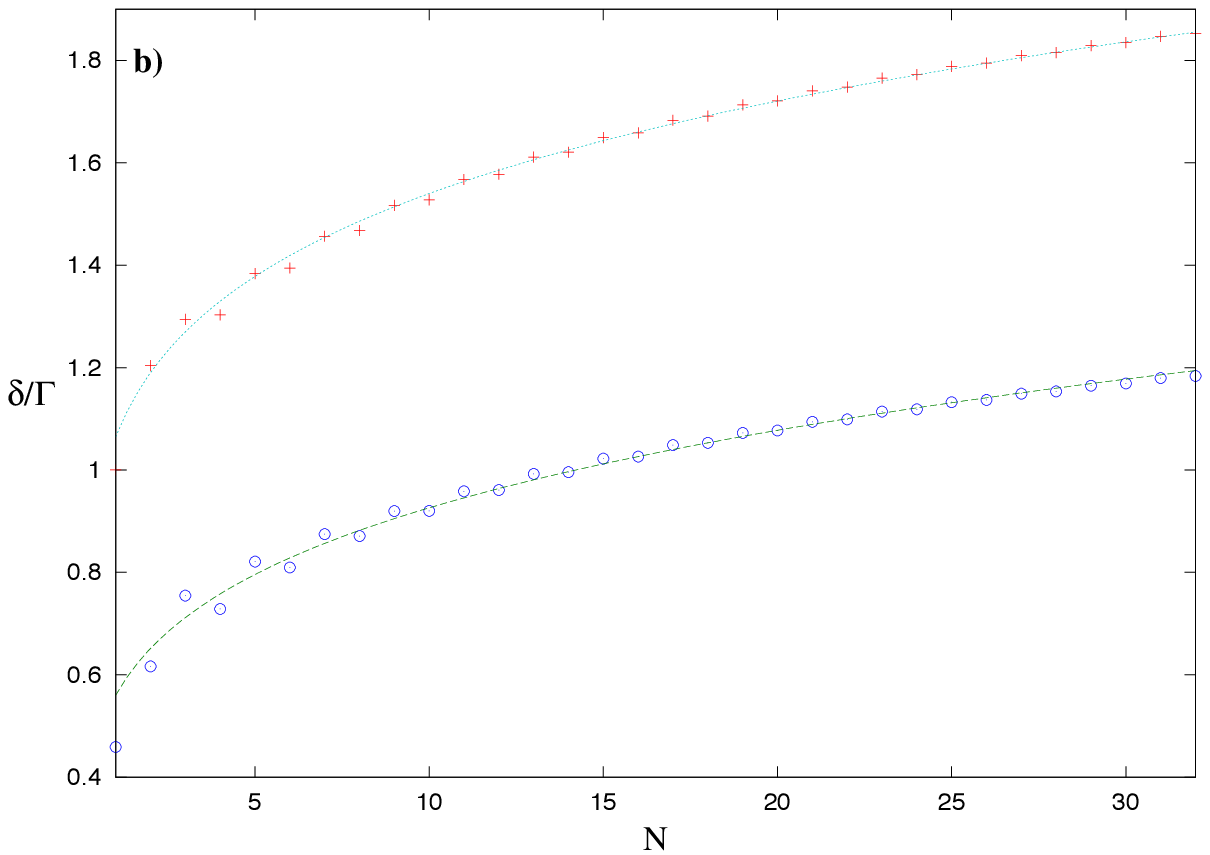}
\caption{(a) Change of the detector efficiency as we introduce a
  detuning between the photon and absorber frequencies. (b) Bandwidth
  of the photo detector vs. the number of absorbers. Given the optimal
  working point $\gamma_{opt}$ which maximizes the detection
  efficiency at zero detuning, $\delta=\omega-\omega_0,$ we look for
  the value of the detuning such that the efficiency at $\gamma =
  \gamma_{opt}(1+1i\delta)$ is reduced by 5\% (circles) or 20\%
  (crosses). The solid lines are fits to $N^{0.16}$ and $N^{0.2}.$}
\label{fig:bandwidth}
\end{figure}

There are two other sources of error which are intimately
connected. One is the bandwidth of the incoming photons, $\Delta
\omega$, which can be related to the length of the photon wavepacket,
$\tau \sim 1/\Delta \omega.$ Since the detector efficiency depends on
$\gamma \propto \Gamma + i (\omega - \omega_0),$ which contains the
detuning between the photon and qubit frequencies, some spectral
components may only be weakly detected. To solve this problem we need
the detector efficiency to be weakly sensitive to variations in the
detuning of the photon, $\delta = \omega - \omega_0,$ that is we need
a broadband detector.

To estimate the bandwidth of the photodetector we have applied the
following criterion. We compute the optimal value of $\gamma_{opt}$
for which the detection happens at maximum efficiency with zero
detuning, $\delta = \omega-\omega_0=0.$ We then look for the two
values of the detuning at which $\gamma =
\gamma_{opt}(1+1i\delta/\Gamma)$ causes a reduction of the efficiency
of a given percentage. As Fig.~\ref{fig:bandwidth}b shows, the
bandwidth grows with a power law $N^\nu,$ with an exponent which is
unfortunately not too large. However, Fig.~\ref{fig:bandwidth}a shows
that the detuning may actually increase the efficiency, probably
indicating that the previous analysis is too limited and that the
detector design may involve optimizing both $\Gamma$ and $\delta$ not
only for achieving a certain efficiency, but also to increase the
efficiency.

Another related problem is dephasing. As we will discuss later on, quantum
circuits are affected by 1/f noise. Part of this noise can be
understood as oscillating impurities that change the electromagnetic
environment of the absorbers, and thus the relative energies of the
$0$ and $1$ states. This error source is modeled with a term $\epsilon
\sigma^z = \epsilon (\ket{1}\bra{1} - \ket{0}\bra{0}),$ where
$\epsilon$ is a random variable ---either classical, or quantum, from
a coupling with the environment---. When one averages over the
different noise realizations, the result is decoherence.

We can get rid of the noise \textit{for each realization} using a
unitary operator $U(t) \sim \exp(-i \int_0^t \epsilon \sigma^z d\tau
/\hbar).$ When we use this operator to simplify the Schr\"odinger
equation, creating the equivalent of an interaction picture, the
result is that we can translate the accumulated random phase,
\begin{equation}
  \phi(t,\epsilon) = \int_0^t \epsilon d\tau / \hbar,
\end{equation}
into the coupling
\begin{equation}
  H_{int} \sim V\int\delta(x)\left[e^{i\phi_\epsilon}\psi_r +
    e^{i\phi_\epsilon}\psi_l\right]
  \ket{1}\bra{0} + \mathrm{H.c.} 
\end{equation}
In other words, from the point of view of the absorber it is as if the
incoming photon was affected by a random, but slowly changing
phase~\footnote{Remember that the noise source is 1/f and dominated by
  low frequencies} that may either shift the frequency of the photons
or broaden their spectral distribution. In either case, if the
photodetector has a large enough bandwidth we may expect just a minor
change in the detector efficiency.

\section{Model implementation: microwave guide with phase qubits}
\label{sec:application}

In this section we detail a possible implementation of our scheme
which is based on elementary circuits found in today's experiments with
superconducting qubits. We will show how our previous theory relates
to the mesoscopic physics of these circuits and compute expressions
for the relevants parameters, $v, V, \gamma, \ldots$ in terms of the
properties of these circuits.

For the waveguide we will consider a coaxial planar microwave guide
such as the ones employed to manipulate and couple different
qubits~\cite{wallraff04,day03} and described in detail in
Sec.~\ref{sec:waveguide}. Note, however, that unlike in
Ref.~\cite{wallraff04} our waveguide will not be cut at the borders
and it will not form a resonating cavity. For the bistable elements we
will consider a superconducting qubit, the so called ``phase qubit''
or ``current-biased Josephson junction'' (CBJJ), which has a set of
metastable levels that, by absorption and emission of photons, may
decay to a different, macroscopically detectable current state.

\subsection{Current-biased Josephson junction}

\begin{figure}[t]
  \centering
  \includegraphics[width=0.7\linewidth]{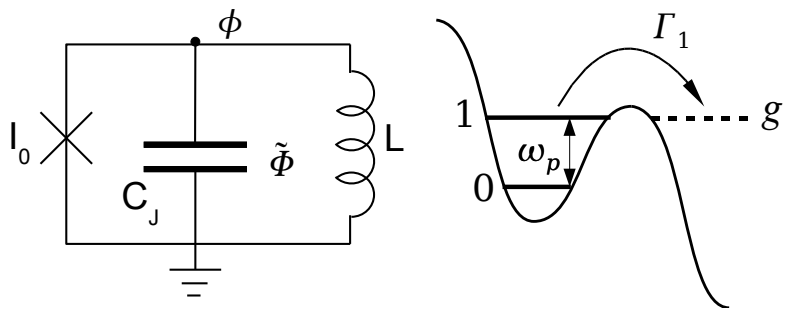}
  \caption{(left) Scheme for a current-biased Josephson
    junction. (right) Energy levels. There are $N_s$ metastable
    energy levels with anharmonic frequencies.}
  \label{fig:cbjj}
\end{figure}

As mentioned before, our detection element will be a CBJJ.  The model
for this circuit is shown in Fig.~\ref{fig:cbjj}: there is a Josephson
junction shunted by an current source which can be modeled by a very
big impedance. The bias current $I$ causes a tilting of the energy
potential in the junction, creating metastable regions with a finite
number of energy levels, that tunnel quantum-mechanically outside the
barrier.

The quantization of this circuit renders a simple Hamiltonian~\cite{devoret04}
\begin{equation}
  H = \frac{1}{2C_J} Q^2 + U(\phi),
\end{equation}
expressed in terms of the charge in the junction and the flux $\phi,$
at a node of the circuit [Fig.~\ref{fig:cbjj}]. The Hamiltonian
contains the usual capacitive energy, expressed in terms of the large
capacitance of the junction, $C_J,$ and a potential energy due to the
inductive elements. Modeling the current source as a large inductor
with a total flux that supplies a constant current, $\tilde\Phi / L_J
= I,$ we obtain a highly anharmonic potential
\begin{equation}
  U(\phi) = - I_0\varphi_0 \cos(\phi/\varphi_0) - I\phi.
  \label{potential}
\end{equation}
Note that even though the actual flux quantum is $\phi_0=h/2e,$ in
order to avoid $2\pi$ factors everywhere it is convenient to work with
$\varphi_0 =\phi_0/2\pi.$

The quantization of this model corresponds to imposing the usual
commutation relations between the canonically conjugate variables, the
flux $\phi$ and the charge $Q$, that is $[\phi, Q] = i \hbar.$ Given
the relevance of the anharmonic terms in Eq.~(\ref{potential}) it soon
becomes evident the convenience of working in the number-phase
representation $\phi = \varphi_0 \times \theta,$ and $Q = 2e \times N$
with operators that satisfy $[\theta, N] = i.$ In these variables the
Hamiltonian becomes
\begin{equation}
  H = E_C N^2 - I_0\varphi_0 \left [\cos(\theta) + \frac{I}{I_0}\theta \right],
\end{equation}
with the junction charging energy
\begin{equation}
  E_C = \frac{(2e)^2}{2C_J} = \frac{\hbar^2}{2\varphi_0^2C_J} \,\, .
\end{equation}

\subsection{Harmonic approximation}

When the bias current $I$ is very close to the critical current $I_0,$
we have the situation in Fig.~\ref{fig:cbjj}, in which the junction
develops a metastable, local minimum of the potential at $\theta$
close to $\pi/2.$ It is then customary to approximate the potential by
a cubic polynomial and describe the dynamics semiclassically, with a coherent
component that describes the short-time oscillations around the local
minimum and a decay rate to the continuum of charge states which are
outside this unstable minimum.

The semiclassical limit is characterized by just two numbers, the plasma
frequency of the phase oscillations around the minimum
\begin{equation}
  \omega_p = \sqrt{\frac{I_0}{4\varphi_0 C_J}}
  \left[1 - \left(\frac{I}{I_0}\right)^2\right]^{1/4}
\end{equation}
and the barrier height
\begin{equation}
  \Delta U = \frac{2\sqrt{2}}3 \sqrt{1 - \frac{I}{I_0}},
\end{equation}
that prevents tunnneling outside this minimum. Using semiclassical
methods it is possible to estimate the number $\sim N_s = \Delta
U/\hbar\omega_p$ of metastable states in this local minimum, and
approximate their energy levels,
\begin{equation}
  E_n/\hbar = n \omega_p + \omega_n^{anh} - i \Gamma_n,
  \label{complex-states}
\end{equation}
including an anharmonic component, $\omega_n^{anh},$ and an imaginary
part which is the rate at which the state decays into the continuum
\begin{eqnarray}
  \Gamma_n = \omega_p \frac{(432 N_s)^{n + 1/2}}{\sqrt{2\pi}n!}
  e^{-36N_s/5}.
  \label{decay-rates}
\end{eqnarray}
In this work, we use the fact that for two consecutive levels these
rates can be very different, $\Gamma_{n+1}/\Gamma_{n}\sim 1000.$ If
the decay rate of the first level $\Gamma_1$ is large enough while
still keeping $\Gamma_0$ small, we can treat the levels $\ket{0}$ and
$\ket{1}$ as the two levels of our absorbing element, the continuum
playing the role of the detectable states $\ket{g}.$ With this we have
now the parameters $\omega=\omega_p$ and $\Gamma\equiv\Gamma_1$ of our
model. It only remains to find out the coupling $V$ and the group
velocity $v.$

In order to quantify the coupling between the junction and the
microwave guide we will need to explicitely make the calculations
leading to the previous results and find out the harmonic
approximations for the lowest energy levels. We begin by noticing that
the energy minima are reached at a value of the phase
\begin{equation}
  \sin(\theta_m) = \frac{I}{I_0} =: r
\end{equation}
which is very close to $\pi/2,$ as predicted. Around this minimum,
$\theta=\theta_m+\tilde\theta,$ a harmonic approximation gives
\begin{equation}
  H \simeq E_C \hat{N}^2 + \frac{1}{2}U'' \tilde\theta^2,
\end{equation}
with a curvature of the potential
\begin{equation}
  U''(\theta_m) = I_0\varphi_0\cos(\theta_m) = I_0\varphi_0\sqrt{1 - r^2}.
\end{equation}
To diagonalize this Hamiltonian we introduce operators
\begin{eqnarray}
  \hat N &=& \frac{1}{\sqrt{2}\alpha}i(a^\dagger-a),\label{charge}\\
  \tilde\theta &=& \frac{\alpha}{\sqrt{2}}(a+a^\dagger),
\end{eqnarray}
with usual commutation relations, $[a,a^\dagger]=1,$ and impose
\begin{equation}
 \frac{E_C}{2\alpha^2} = \frac{U''\alpha^2}{4} = \frac{1}{2}\hbar\omega_p \, .
\end{equation}
This provides us with the plasma frequency,
\begin{eqnarray}
 \omega_p &=& \sqrt{\frac{E_CU''}{2\hbar^2}} = \sqrt{\frac{I_0}{4\varphi_0C_J}}
  (1-r^2)^{1/4},
\end{eqnarray}
but also with the ``strength'' of the number and phase operators,
\begin{equation}
 \alpha^2 = \sqrt{\frac{2E_C}{U''}} = \frac{E_C}{\hbar\omega_p} \, ,
 \label{alpha}
\end{equation}
which is related to the charging energy and the plasma frequency
in a simple way.

\subsection{Qubit-line in the dipole approximation}

Given that the total capacitance of the transmission line is much
larger than that of a single superconducting qubit, we can apply the
dipole limit to study the coupling between both elements and assume
that the interaction term goes as
\begin{equation}
H = \hat{q} \frac{C_g}{C_g + C_J} \hat{V}(x,t).
\label{dipole-coupling}
\end{equation}
Following the theory in Sec.~\ref{sec:waveguide}, the potential
created inside the waveguide can be written as a function of the
charge distribution on the conductor
\begin{equation}
  \hat{V}(x,t)=\frac{1}{c}\frac{\partial \hat{Q}(x,t)}{\partial x} \, .
\end{equation}
Using the eigenvalue equation $\partial_x w_k(x,t) = i k w_k(x,t)$ and
the dispertion relation (\ref{dispersion}), we obtain
\begin{equation}
  \hat{V}(x,t)=i\sum_k \sqrt{\frac{\hbar \omega_k}{2
      c}}\left[a_k w_k(x,t) - a_k^\dagger w_k(x,t)^*\right].
\end{equation}
If we assume that the incident and outgoing wavepackets have a small
bandwidth, $\omega_k \simeq \omega_{0},$ we obtain the potential as a
function of the propagating fields
\begin{equation}
  \label{field-coupling}
  \hat{V} \simeq i \sqrt{\frac{\hbar\omega_{0}}{2c}}
  \left[\psi_r + \psi_l - \psi_r^\dagger - \psi_l^\dagger\right].
\end{equation}
As a consistency check, we can consider the case of a small
transmission line, forming a ``cavity'' or resonantor in which the
modes are very well separated. In that limit we only need to consider
a single momentum, $p,$ and everything reduces to the formula
\begin{equation}
\hat{V}(x,0)=i\sqrt{\frac{\hbar \omega_p}{2 C}}(a_p-a^{\dag}_p) \, .
\end{equation}
Here, $C=L\times c$ is the total capacity of the transmission line and
the quantity $(\hbar \omega_k/2C)^{1/2}$ is the r.m.s. voltage. Note
that, consistently with the expression in the experiment of Wallraff
\textit{et al.}~\cite{blais04,wallraff04}, the coupling between the
qubit and a particular mode is inversely proportional to the square
root of the microwave guide length, $L.$ However, the coupling
constant with the fields in Eq.~(\ref{field-coupling}) is insensitive
to the total size of the transmission line.

\subsection{Final parameterization of the setup}

Following the previous considerations we will write the interaction
Hamiltonian between the junction and the microwave field as in
Eq.~(\ref{dipole-coupling}) using the charge operator
(\ref{charge}). The coupling between the stripline and the two lowest
levels of the CBJJ is proportional to the constant
\begin{equation}
  V = \frac{C_g}{C_J+C_g} \frac{e}{\alpha}
  \sqrt{\frac{\hbar\omega_0}{c}}.
\end{equation}
Let us remind that $\omega_0$ is the photon frequency, $C_g$ is the
gate capacitance between the junction and the waveguide, $c$ is the
capacitance per unit length of the waveguide and $\alpha$ is the
harmonic oscillator wavepacket size (\ref{alpha}).

In addition to the coupling strength we need to compute the group
velocity $v.$ The precise derivation is left for
Sec.~\ref{sec:waveguide}, but we advance that this can be
written using properties of the waveguide
\begin{equation}
  v = \frac{1}{\sqrt{cl}} = \frac{1}{cZ_0},
\end{equation}
where $l$ is the inductance per unit length and $Z_0$ the impedance.
When introduced in our scattering problem, this results in the
effective decay rate $\gamma$
\begin{equation}
  \gamma = \frac{\hbar v}{V^2} \hbar \Gamma
  = \frac{\hbar/cZ_0}{c_{12}^2e^2\hbar\omega_{0}/\alpha^2 c}
  \hbar \Gamma
  = \frac{\alpha^2}{c_{12}^2}
  \frac{\hbar}{e^2Z_0} \frac{\Gamma}{\omega_{0}},
\end{equation}
expressed in terms of $c_{12} = C_g/(C_g + C_J)$ for resonant qubits
(i.e. no detuning). We want to remark that there are plenty of
parameters to tune, so that realistic conditions of high efficiency
are achievable.

In practice one would begin by fixing the number of superconducting
absorbers that can be used, $N$. This fixes a value of $\gamma$ for
which absorption is optimal and which determines the optimal
decay rate of the first energy level
\begin{equation}
  \Gamma = \frac{c_{12}^2}{\alpha^2} \frac{Z_0e^2}{h} \omega_0 \, .
\end{equation}

As an example, let us compute the optimal configuration for one qubit.
Using the numbers in Berkley's experiment~\cite{berkley03b}, we will
fix the qubit capacitance $C_J=4.8$pF and $c_{12}=0.13.$ The charging
energy is about $E_C/\hbar = 0.10$ GHz, which together with a plasma
frequency $\omega = 5$ GHz gives $\alpha^2=0.02.$ Putting the numbers
together
\begin{equation}
  \Gamma = \frac{Z_0}{5144\Omega} \omega_0,
\end{equation}
so that for inductances from $10\Omega$ to $100\Omega$ this gives
$\Gamma=10-100$ MHz. Higher decay rates can be achieved by increasing
$C_g$ or decreasing the charging energy. For instance, a factor 2
increase of $C_g$ gives a factor 4 increase in $\Gamma$ which now
ranges from $30$-$300$~MHz.

\section{Photons in transmission lines}
\label{sec:waveguide}

In this section we study the mathematical models and physical
properties of a coaxial coplanar microwave transmission line. This
device consists on a central conductor carrying the wave and enclosed
by two conductors set to the ground plane. A simple but effective
model for such a line is obtained by coupling inductances and
capacitors as shown in Fig.~\ref{fig:schemeTL}. As we will show later
on, if we denote by $l$ and $c$ the inductances and capacitances per
unit length, this model predicts the propagation of microwave fields
in the transmission line according to a dispersion relation
\begin{equation}
  \omega_k = v |k| = \frac{1}{\sqrt{cl}} |k|,
  \label{dispersion}
\end{equation}
where $\hbar k$ is the momentum of the photon, $\omega_k$ is the
frequency and $v$ is the group velocity.

\subsection{Discrete model}

\begin{figure}[t]
  \centering
  \includegraphics[width=0.6\linewidth]{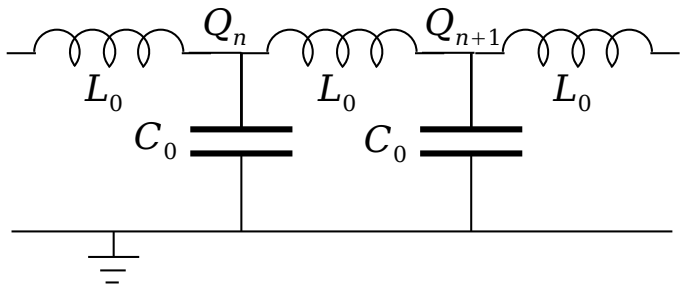}
  \caption{A transmission line can be modeled as a set of inductances
    and capacitances.}
  \label{fig:schemeTL}
\end{figure}

In order to analyze this circuit from Fig.~\ref{fig:schemeTL}, we
must write down the Kirchhoff's law for the n-th block containing 4 nodes. When combining all equations and leaving
as only variables the branch intensities, we obtain the set of
second order differential equations
\begin{equation}
-L_0\frac{d^2I_{n}}{dt^2}=\frac{1}{C_0}[2I_n-I_{n-1}-I_{n+1}].
\label{seconddiff}
\end{equation}

These equations are similar to those describing an infinite set of
oscillators of mass $m\propto L_0$ and spring constant $\kappa\propto
1/C_0.$ In analogy with the mechanical case, if we assume periodic
boundary conditions to better reproduce propagation of charge, we find
travelling wave solutions
\begin{equation}
I_n(t)=I_0e^{i(kx_n-\omega t)},\; k = \frac{2\pi}{L}\times \bf{Z},
\end{equation}
where $x_n=a\times n,$ $a$ is a parameter denoting the distance
between neighbor oscillators, $p$ is the momentum of the wave and $L$
is the length of the line. A direct substitution of this expression in
Eq.~(\ref{seconddiff}), gives the dispertion relation
\begin{equation}
\omega(k)=\left[\frac{2}{L_0C_0}\left(1-\cos(ka)\right)\right]^{\frac{1}{2}}
\simeq \sqrt{\frac{a^2}{L_0C_0}} |k|,
\end{equation}
which is approximately linear for small momenta, long waveguides or
thin discretization. Using the inductance and capacitance per unit
length
\begin{equation}
  l = L_0/a,\; c=C_0/a,
\end{equation}
we obtain the group velocity $v$ and dispersion relations introduced
before~(\ref{dispersion}).

\subsection{Lagrangian formalism and continuum limit}

The previous evolution equations~(\ref{seconddiff}) can be obtained
from the Lagrangian
\begin{equation}
L=\sum_n\left(\frac{L_0}{2}\dot{Q}^2_n-\frac{1}{2C_0}(Q_n-Q_{n+1})^2\right)
\end{equation}
using the Euler-Lagrange equations
\begin{equation}
  \frac{d}{dt}\left(\frac{\partial L}{\partial\dot{Q}_n}\right)=
  \frac{\partial L}{\partial Q_n} \, .
\end{equation}
A more realistic model for a continous transmission line is obtained
by taking the limit of infinitesimally small capacitors and inductors,
$a \! \to \! 0,$ while preserving the intensive values $l$ and $c$. In the
continuum limit we replace the discrete charges by a continuous charge
density distribution, $Q_n \simeq a \times q(x_n,t),$ which should be
a smooth and integrable function of the position variable $x$. Through
this procedure one obtains a total Lagrangian expressed in integral
form
\begin{equation}
L=\int dx \left[\frac{l}{2}(\partial_tq)^2-
  \frac{1}{2c}(\partial_x q)^2\right].
\end{equation}
The charge distribution described through the field $q(x,t)$ obeys a
wave equation
\begin{equation}
\partial^2_xq(x,t)-\frac{1}{v^2}\partial^2_tq(x,t)=0,
\label{waveeq}
\end{equation}
where the group velocity is precisely the one already found in
the discrete case.

\subsection{Hamiltonian and quantization}

The quantization of an electrical circuit is a three-step process. We
begin by identifying the classical canonically conjugate variables:
the charge distribution $q(x,t)$ and the associated momentum
\begin{equation}
  \Pi_q = \frac{\partial {\cal L}}{\partial \dot q} = l \dot q(x,t).
  \label{momentum}
\end{equation}
The Hamiltonian is then built using the prescription
\begin{equation}
  H=\int dx \Pi_q\dot{q} - L=
  \int dx \left[
  \frac{(\Pi_q)^2}{2l}+\frac{(\partial_x q)^2}{2c}\right].
  \label{hamilt}
\end{equation}
Finally, the variables $\{q,\Pi_q\}$, are replaced with Hermitian
operators $\{\hat{q}, \hat{\Pi}_q\},$ with commutation relations
\begin{equation}
\big[\hat{q}(x,t),\hat{\Pi}_q(x^{\prime},t)]=i\hbar\delta(x-x^{\prime}).\label{conmut}
\end{equation}

The previous Hamiltonian is quadratic and it can be diagonalized. We
begin by analyzing the dynamics of these operators in the Heisenberg
picture and noticing that the charge satisfies a wave equation
(\ref{waveeq}). It therefore makes sense to look for a set of normal
modes made of plane waves
\begin{eqnarray}
\hat{q}(x,t) & = & \sum_k N_k \hat{q}_k(t) u_k(x) , \quad \\
u_k(x,t) & = & \frac{1}{\sqrt{{\cal L}}} \exp(ikx),\;
k \in \frac{2\pi}{{\cal L}}\times \bf{Z},
\end{eqnarray}
where $\hat{q}_k$ are our new dynamical variables and $N_k$ is a
normalization constant which will be fixed later on. Note that we are
using periodic boundary conditions because they are best suited for
describing transport, but we have not fixed the transmission line
length ${\cal L},$ which may be arbitrarily large.

Replacing the expansion above in~(\ref{waveeq}), produces second
order differential equation for the unknown operators
\begin{equation}
\ddot{\hat{q}}_k(t)+\omega^2_k\hat{q}_k(t)=0.
\end{equation}
Since both $\hat{q}$ and $\hat{q}_k$ are physical observables and
Hermitian operators, the most general solution is
$\hat{q}_k(t)=a_k e^{-i\omega_kt}+a_k^{\dagger}e^{i\omega_kt},$
expressed in terms of a time independent operator, $a_k.$ The
commutation relation~(\ref{conmut}) produces both bosonic commutation
relations for the $a_k$
\begin{equation}
  [a_k,a^{\dag}_{k'}]=\delta_{kk'},\; [a_k,a_{k'}] = [a_k^\dagger,a_{k'}^\dagger]=0,
\end{equation}
and $N_k^2 = \hbar/(2\omega_k l).$ Finally, with the orthonormalized
wave functions $w_k(x,t) = u_k(x) \exp(-i\omega_k t),$ we arrive at
\begin{eqnarray}
  \hat{q}(x,t)&=&\sum_k\,\sqrt{\frac{\hbar}{2\omega_kl}}
  \left[a_k w_k(x,t)+\mathrm{H.c.}\right]\label{chargeP}
  \\
  \hat{\Pi}_q(x,t)&=&\sum_k\,i\sqrt{\frac{\hbar
      \omega_k l}{2}}\left[a^{\dag}_kw_k(x,t)^*-\mathrm{H.c.}\right].\nonumber
\end{eqnarray}

Given the specific form of the canonical operators, we may obtain a
particular dispersion relation $\omega_k$ that diagonalizes the
Hamiltonian. Using the relations
\begin{eqnarray}
  \int w_k(x,t) w_{k'}(x,t) dx &=& \delta_{k + {k'}}e^{-i(\omega_k + \omega_{k'})t},\\
  \int w_k(x,t)w_{k'}(x,t)^* dx &=& \delta_{k - {k'}},
\end{eqnarray}
and imposing
\begin{equation}
  \frac{\hbar\omega_kl}{2}\times\frac{1}{2l} = \frac{\hbar}{2\omega_kl}
  \times\frac{1}{2c} \times k^2,
\end{equation}
we will be able to cancel all terms proportional to $a_ka_{-k}$ and
$a_{k}^\dagger a_{-k}^\dagger,$ obtaining a set of uncoupled
oscillators
\begin{equation}
  H = \sum_k \hbar\omega_k \left(a^\dagger_k a_k +  \frac{1}{2}\right),
\end{equation}
where the dispersion relation is strictly the one introduced before in
Eq.~(\ref{dispersion}).

\subsection{Linearization}

In our work we focus on states that contain photons with momenta
around $|k_0|$ or $\omega_{0}/v,$ where $\omega_{0}$ is the principal
frequency of the wavepacket. We thus introduce two field operators
representing the right- and leftward propagating photons,
\begin{eqnarray}
  \psi_r(x,t) &=& \sum_{k\in{\cal B}}a_k w_k(x,t),\\
  \psi_l(x,t) &=& \sum_{k\in{\cal B}}a_{-k} w_{-k}(x,t),
\end{eqnarray}
where ${\cal B} = [k_0-\Delta,k_0 + \Delta]$ is the desired
neighborhood around the principal momentum, characterized by a
sensible cut-off $\Delta.$ These two fields satisfy the evolution
equations
\begin{eqnarray}
  i\partial_t \psi_r(x,t) &=& -i \hbar v\partial_x\psi_r(x,t),\\
  i\partial_t \psi_l(x,t) &=& +i \hbar v\partial_x\psi_l(x,t),
\end{eqnarray}
and we can write the effective Hamiltonian
\begin{equation}
  H = \int \left[\psi_r^\dagger(-i\hbar v\partial_x)\psi_r +
    \psi_l^\dagger(+i\hbar v\partial_x)\psi_l\right]dx.
\end{equation}

There is a catch in the previous expansion: the operators $\psi_{r,l}$
do not have the usual bosonic commutation relations. For instance
\begin{equation}
  [\psi_{r}(x,t),\psi_{r}^\dagger(y,t)] =
  \sum_{k\in{\cal B}} w_{k}(x,t){\bar w}_{k}(y,t),
\end{equation}
where the right hand side represents a truncation of the distribution
$\delta(x-y)$ within the interval of momental ${\cal B}.$ In this work
we are interested in the asymptotics of the absorption process,
treating long times and long enough wavepackets for which the rotating
wave approximation is valid. In this case we are justified to
approximate
\begin{equation}
  [\psi_{\alpha}(x,t),\psi_{\beta}^\dagger(y,t)] \simeq
  \delta_{\alpha,\beta}\delta(x-y),~~ \alpha,\beta\in\{r,l\},
\end{equation}
and treat the right and left propagating fields as causal.

\section{Conclusions}
\label{sec:final}

We have developed the theory of a possible microwave photon detector in circuit QED, and studied diverse regimes, advantages and difficulties with a realistic scope. Though we believe that our contribution will boost the theoretical interest and possible implementations in microwave photodetecion, we will summarize the potential limitations and imperfections of our proposal. First, the bandwidth of the detected photons has to be small compared to the time required to absorb a photon, roughly proportional to $1/\Gamma.$ Second, the efficiency might be limited by
errors in the discrimination of the state $\ket{g}$ but these effects
are currently negligible~\cite{hofheinz08}. Third, dark counts due to
the decay of the state $\ket{0}$ can be corrected by calibrating
$\Gamma_0$ and postprocessing the measurement statistics. Fourth,
fluctuations in the relative energies of states $\ket{0}$ and
$\ket{1},$ also called dephasing, are mathematically equivalent to an
enlargement of the incoming signal bandwidth by a few megahertz and
should be taken into account in the choice of parameters. Finally, and
most important, unknown many-body effects cause the non-radiative
decay process $1\to0,$ which may manifest in the loss of photons while
they are being absorbed. In current experiments~\cite{hofheinz08}, this
happens with a rate of a few megahertzs, so that it would only affect long wavepackets.

Our design can be naturally extended to implement a photon counter
using a number of detectors large enough to capture all incoming
photons.  Furthermore, our proposal can be generalized to other level
schemes and quantum circuits that can absorb photons and irreversible
decay into long lived and easily detectable states.

We expect to have contributed to the emerging field of detection of travelling photons. Its success may open the doors to the arrival of ``all-optical'' quantum information processing with propagating quantum microwaves.

\section*{Acknowledgments}

The authors thank useful feedback from P. Bertet, P. Delsing,
D. Esteve, M. Hofheinz, J. Martinis, G. Johansson, M. Mariantoni,
V. Shumeiko, D. Vion, F. Wilhelm and C. Wilson. G.R. acknowledges
financial support from CONICYT grants and PBCT-Red 21, and hospitality
from Univ. del Pa\'{\i}s Vasco and Univ. Complutense de Madrid. J.J.G.-R. received support from Spanish Ramon y Cajal program, and projects FIS2006-04885 and CAM-UCM/910758. E.S. thanks support from Ikerbasque Foundation, UPV-EHU Grant GIU07/40, and EU project EuroSQIP.

%\bibliographystyle{Harvard}
%\bibliography{prb}

\end{document}